\documentclass[preprint,12pt]{elsarticle}

\usepackage{amssymb}
\usepackage{graphicx}
\usepackage{epsfig}
\usepackage[small]{caption}
\usepackage{lineno}

\linenumbers

\journal{Nuclear Instruments and Methods A}

\begin{document}

\begin{frontmatter}


\title{A Xenon Condenser with a Remote Liquid Storage Vessel}

\author[umd]{S.~Slutsky
\footnote{Corresponding author. Please submit correspondence via e-mail to simons@umd.edu.} }
\author[umd]{Y.-R.~Yen}
\author[umd]{H.~Breuer}
\author[umd]{A.~Dobi}
\author[umd]{C.~Hall}
\author[umd]{T.~Langford}
\author[umd]{D.~S.~Leonard}
\author[umd]{L.~J.~Kaufman}
\author[triumf]{V.~Strickland}
\author[umd]{N.~Voskanian}

\address[umd]{Department of Physics, University of Maryland, College Park, MD 20742 USA}
\address[triumf]{TRIUMF, 4004 Wesbrook Mall, Vancouver, BC, V6T 2A3 Canada \\
P. Eng. at Carleton University, Ottawa, Canada }

\begin{abstract}

We describe the design and operation of a system for xenon liquefaction in which the
condenser is separated from the liquid storage vessel.  The condenser is cooled by 
a pulse tube cryocooler, while the vessel is cooled only by the liquid xenon itself. This
arrangement facilitates liquid particle detector research by allowing easy access to the 
upper and lower flanges of the vessel. We find that an external xenon gas pump is useful 
for increasing the rate at which cooling power is delivered to the vessel, and we present 
measurements of the power and efficiency of the apparatus.

\end{abstract}


\end{frontmatter}

\section{Introduction}
\label{}

Particle detectors based on condensed noble gases have found wide
application in high energy physics, astro-particle physics, and nuclear physics.
Noble liquids are attractive candidates for particle detectors due to their
ease of purification, good charge transport properties, high scintillation efficiency,
and in the case of xenon, high density and short radiation length. Examples of
recent rare-event searches based on cryogenic noble gases include
CLEAN/DEAP \cite{cleandeap}, XMASS \cite{xmass}, ZEPLIN \cite{zeplin},
XENON \cite{xenon}, LUX \cite{lux}, WARP \cite{warp}, ArDM \cite{ardm},
EXO \cite{exo200}, ICARUS \cite{icarus}, and MEG \cite{meg}.

The experimental methods of liquid noble gas detectors have been developed in  
small prototype instruments whose liquid volumes range from a few cubic  
centimeters to tens of liters. In these prototypes, the system is typically cooled  
either by liquid nitrogen or by directly coupling the 
storage vessel to the cold head of a refrigerator. These techniques have several  
attractive features, including simplicity, robustness, and the availability of great  
cooling power. However, both of these cooling strategies typically require a
significant amount of space above or below the storage vessel be devoted to the
cryogenics. In contrast, room temperature detector technologies, such as gas
proportional counters or plastic scintillator, are not burdened by a cryogenic 
system, which is particularly advantageous for prototyping work where the 
freedom to make maximal use of the space around the detector provides valuable 
flexibility.  

In this article we describe a system for condensing and storing a noble  
gas (xenon) where the cooling system is spatially separate from the liquid storage  
vessel, leaving only the cryostat in the space surrounding the vessel.  
This configuration facilitates many common laboratory operations, particularly
those which require
access to the upper or lower flanges of the storage vessel. Examples include
the installation of detector structures and the insertion and removal of material
samples and radioactive sources. This setup also provides direct optical
access to the interior of the vessel for viewing or for laser injection, and it
simplifies the construction of a lead shield.

Our primary motivation for pursuing the remote cooling method discussed
here is to allow the space above the liquid xenon vessel to be used for
ion tagging and retrieval experiments in the context of the EXO double beta
decay search. EXO has proposed to eliminate radioactive backgrounds by
identifying the barium ion produced in the double beta decay of
$^{136}$Xe \cite{danilov}. The identification method may require that a device
be inserted into the active volume of the double beta decay detector to
retrieve the final state nucleus. The condenser and liquid xenon vessel
described in this article will allow both a barium ion calibration source and an
insertion and retrieval device to be coupled to the xenon vessel from above.
 
In the last decade pulse tube cryocoolers have attracted attention as convenient 
and reliable means to liquefy noble gases.
For example, technology has been developed for the MEG and XENON 
experiments using a modified cryocooler integrated into a liquid xenon storage 
vessel \cite{haruyama}. Several other recent articles have reported on the 
development of a small-scale helium condenser based on a pulse tube cryocooler where
the condenser is located directly in the neck
of a liquid helium storage dewar \cite{wang2001}\cite{wang2005}\cite{wangandscurlock}.
In our system, we also employ a pulse tube cryocooler, 
and our condenser is conceptually similar to the helium condenser described in 
references \cite{wang2001}-\cite{wangandscurlock}, although in our case 
the remote storage vessel 
represents an additional complication. Note that we report here the
results of cooling and liquefaction tests carried out with xenon.  Similar results
could likely be obtained with other heavy noble gases having lower 
saturation temperature,
such as argon or krypton, provided that the heat leaks in the system are minimized.  

\section{Apparatus}

\begin{figure}[t!]\centering
\includegraphics[width = 90 mm]{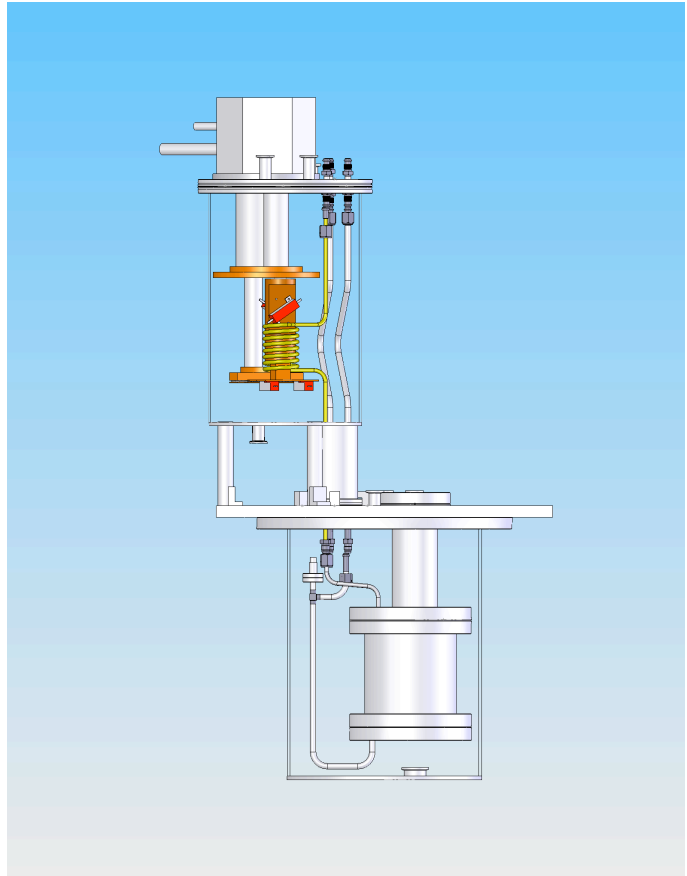}
\caption{A drawing of the full xenon condenser system.  The upper part of
the cryostat holds the condenser, and the lower part holds the liquid xenon
vessel.}
\label{fig:System}
\end{figure}

\begin{figure}[t!]\centering
\includegraphics[width = 90 mm]{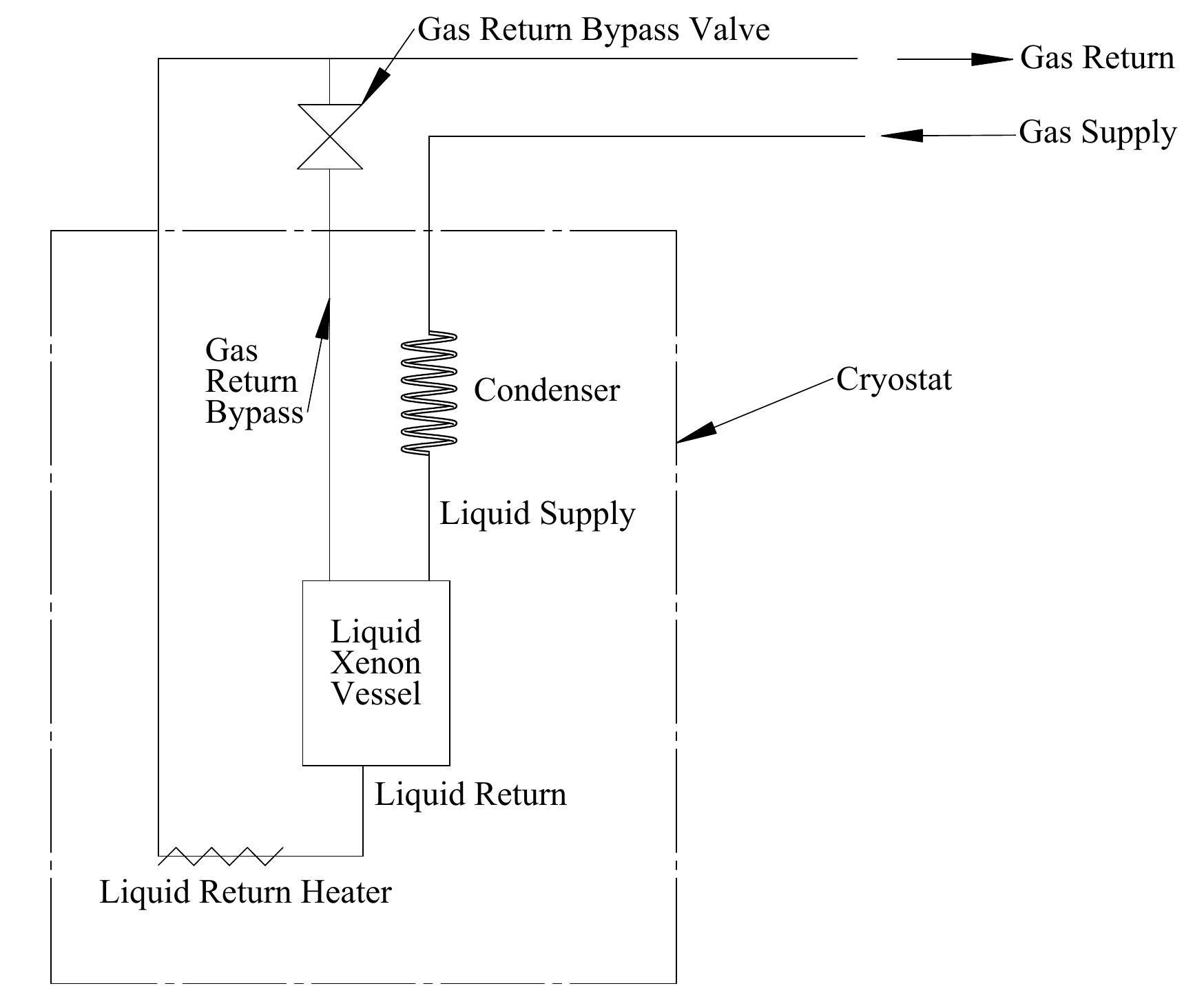}
\caption{A schematic of the condenser system. Room temperature gas flows in
the supply to the condenser.  Liquid drips into the vessel, and cool gas
returns through the gas return line.  The gas can also return through the
liquid return line when no liquid is present.}
\label{fig:Schematic}
\end{figure}

The system consists of two units: a helical copper condenser and a stainless
steel liquid xenon (LXe) storage vessel.  A system drawing and plumbing schematic
are shown in Figures~\ref{fig:System} and \ref{fig:Schematic}, respectively. 
The condenser is located above the vessel, and condensed liquid flows
downward through a 1/4" stainless steel (SS) tube to the top of the vessel. The 
condenser is cooled by the cold head of a cryocooler, while the vessel is  
cooled only by the xenon. The
entire system is enclosed in a vacuum-insulated cryostat composed of two SS
cylinders connected by a bellows. The two cylinders and bellows form one
vacuum volume for pump-out purposes. The radiative heat leak is reduced by
wrapping the system components in super-insulation consisting of 10-15
alternating layers of aluminized mylar and fabric.

\subsection{Condenser and Temperature Control}

The condenser is a helical coil of 1/4" diameter Oxygen Free High Conductivity
copper (OFHC) tube, which was chosen for its purity and thermal conductivity. 
The tubing is partially annealed; non-annealed tube was found to kink
when coiled.  The coil is brazed to a 2" diameter cylindrical shank of OFHC
copper, which is mechanically attached to the coldhead of a pulse tube
cryocooler \cite{cryomech}. The total length of the cooled portion of the
tube is 30 inches (five turns).
   
\begin{figure}[t!]\centering
\includegraphics[width = 90 mm]{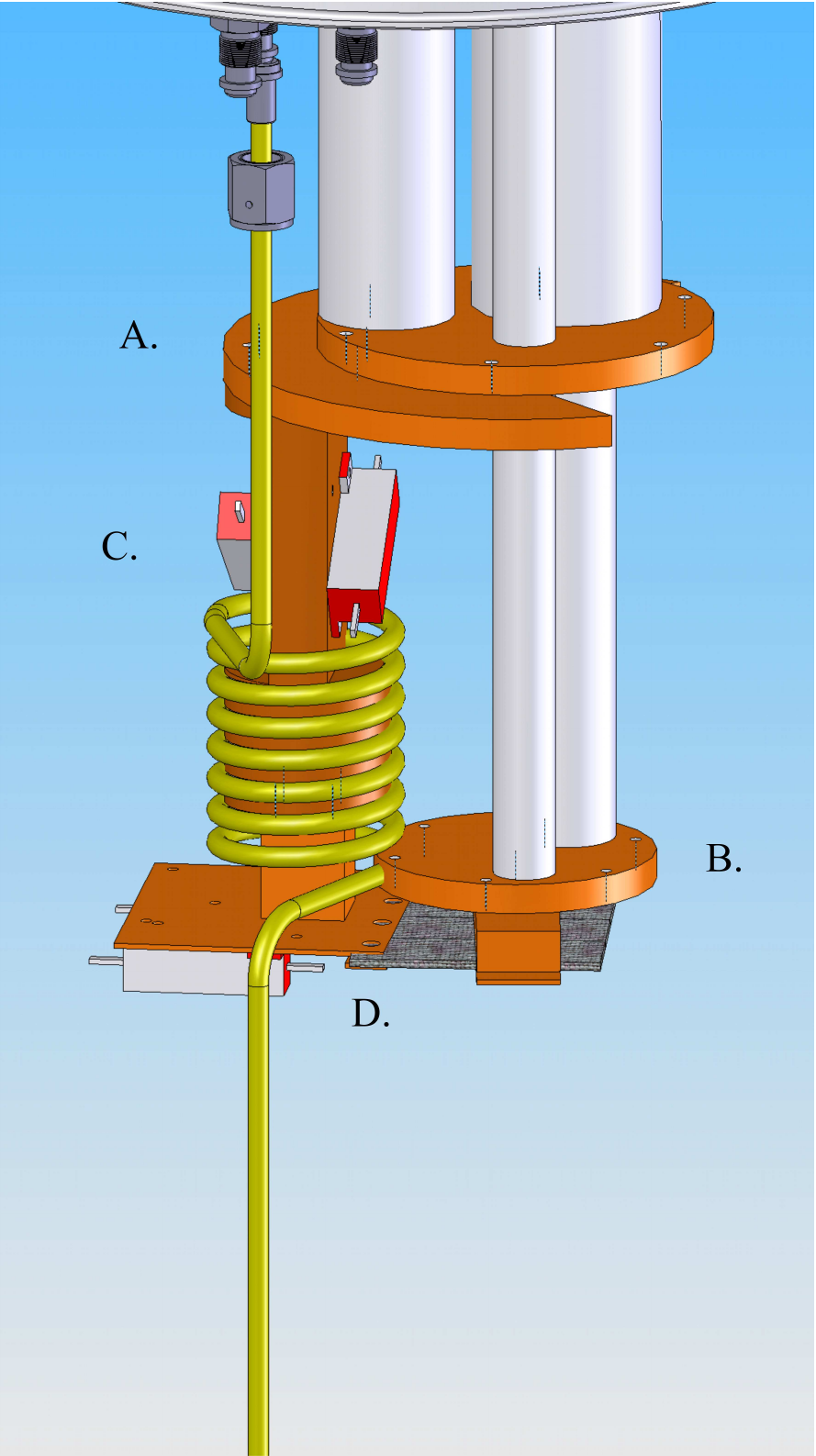}
\caption{A drawing of the condenser. It is coupled to the cryocooler
cold head at both the first stage (A) and second stage (B).
The temperature at each coupling is controlled by trim
heaters (C) and (D).}
\label{fig:Condenser}
\end{figure}


As shown in Figure~\ref{fig:Condenser}, the cold head has two stages, with base
temperatures of 22~K and 8~K for the first and second stages,
respectively.  These base temperatures are far below what is required to
liquefy xenon (triple point of 161~K), but should allow the system to condense the
lighter noble gases.  To ensure temperature uniformity in the condenser, and to
achieve good thermal control, each end of the condenser
is coupled to one of the two stages of the cold head. We refer to the upper half
of the condenser (coupled to the first stage of the cold head) as the ``pre-cooler",
and the lower half (second stage) as the ``post-cooler". The function of the
pre-cooler is to cool the room temperature xenon gas
to the saturation temperature, while the function of the post-cooler is to remove
the latent heat of vaporization, thereby effecting the phase change.

The pre-cooler and the post-cooler are independently temperature controlled  
by trim heaters. The heaters are driven by a common 77~W regulated DC  
power supply, and each heater circuit is controlled by a PID temperature 
feedback unit \cite{omega}. The controllers adjust 
the current in each heater via the gate voltage on two power FETs. 

The pre-cooler trim heaters are mounted on
flats in the cylindrical shank of the condenser. These flats make a narrow ``neck" between
the coil and the upper cold stage, thus reducing the cooling power and allowing for more
precise temperature control.  A stainless steel shim between the condenser shank and the cold head
further reduces the cooling power of the pre-cooler, which was found to be excessive for
our purposes. The post-cooler trim heaters are attached to a plate on the bottom of the
shank, which is itself attached to the second stage of the cold head via a flexible copper
braid. The braid was chosen to provide a flexible thermal bridge so that the condenser is
mechanically constrained at its upper end only. The control temperatures for the two PID
feedback loops are measured by thermocouples located on the neck and on the plate for the  
pre-cooler and post-cooler, respectively.

In typical operation, the pre-cooler set point temperature is chosen to be 179.5~K,
and the post-cooler temperature is chosen to be a few degrees cooler.  In practice,
however, the post-cooler temperature tends to stabilize around 193~K during condensation,
and thus the post-cooler heater does not power on.\footnote{Nevertheless, the post-cooler was found to be crucial for 
maintaining a uniform temperature throughout the condenser; see Section~\ref{sec:discussion}.}
This reflects the fact that the thermal coupling between the post-cooler and the second 
stage of the cold head has too much thermal resistance.  It is likely that this resistance limits the 
power and efficiency of the condenser, so we intend to modify this arrangement in the near future.
Currently, the condensation that does occur is sufficient for our purposes.

The trim heaters on the pre-cooler allow the condenser to adjust for the effects
of changes in the gas flow rate. At high flow rates, significant heat is delivered to
the condenser by incoming room temperature xenon gas, so the pre-cooler trim heater reduces
its power output to maintain the set point temperature.  In some cases, the incoming heat
was found to be sufficient to warm the condenser to one or two degrees above the PID set point.
At zero flow or low flow rates,
the heat delivered to the condenser by the gas is small, and the pre-cooler trim heater  
provides compensation to keep the condenser temperature above the freezing point of xenon. 
Consequently, one benefit of the pre-cooler PID  
feedback loop is that it prevents over-cooling in the event that a large gas  
flow suddenly stops.

We present some measurements of the cooling power of the second stage of the
PT805 near LXe temperature. 
(Cryomech, the cryocooler manufacturer, does not report this data at such high temperatures.)
We attached a known thermal resistance to the second stage of the cold head, applied heat  
from the opposite end, and measured the temperature at each end. Once thermal equilibrium is
reached, the cooling power is equal to the power delivered by the heater, and can
also be inferred from the temperature gradient across the known thermal resistance.
These two methods give results in agreement with each other.  We find that the
cooling power is 25~W at 65~K, 26.2~W at 75~K, and 32.2~W at 152~K.

\subsection{Xenon Vessel}

\begin{figure}[t!]\centering
\includegraphics[width = 90 mm]{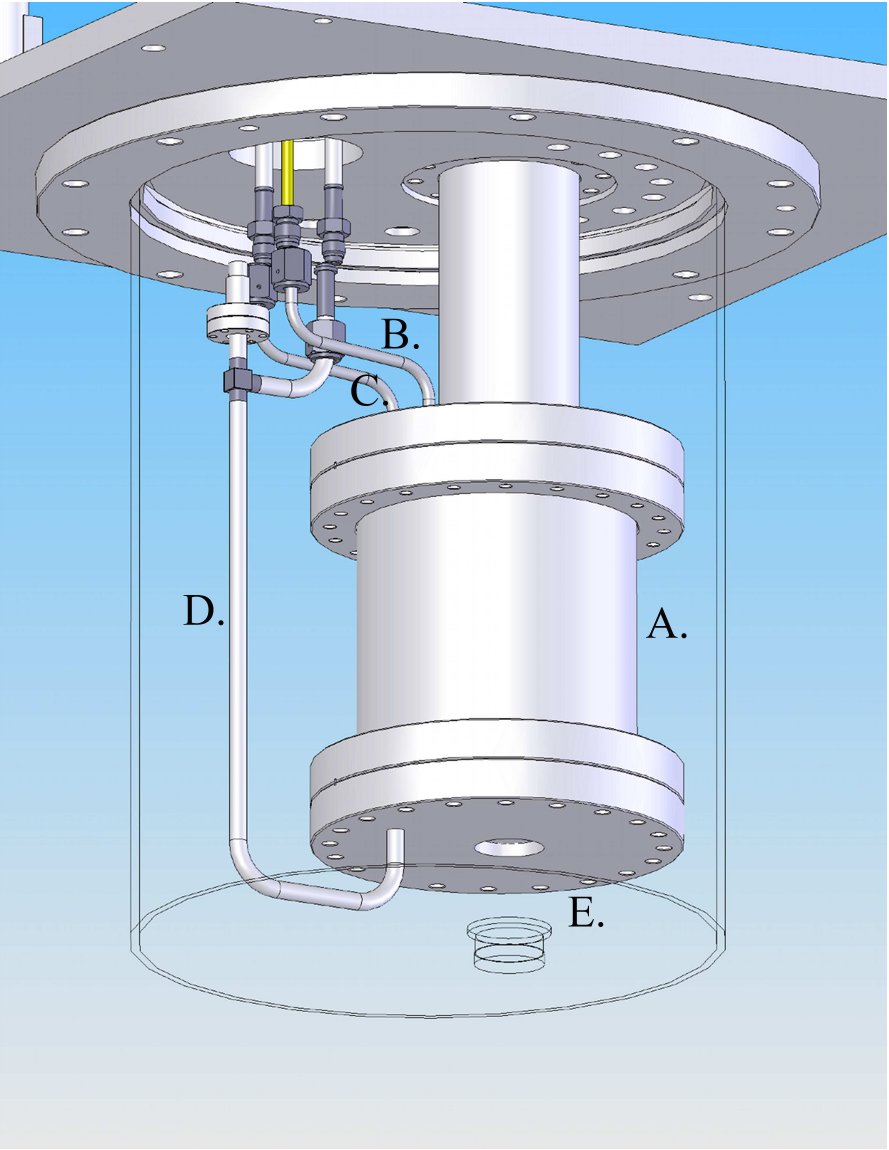}
\caption{A drawing of the xenon vessel (A) and its associated plumbing.  The liquid
supply line (B) and the gas return line (C) can be seen at the top of the vessel. 
The liquid
return line (D) connects to the bottom of the vessel and contains the capacitance level
meter.  The transparent outer cylinder represents the  
cryostat, and the laser injection windows (E) are visible at the  
bottom of the vessel and cryostat. The vessel is supported by the large diameter
(3") tube which penetrates its top flange and permits access to the vessel interior
from the lab.}
\label{fig:Vessel}
\end{figure}


The xenon vessel, seen in Figure~\ref{fig:Vessel}, is a 6" x 6" OD cylinder constructed
from stainless steel, which was chosen for its purity and thermal mass.  This volume
is sufficient to hold 10~kg, or about 3 liters, of LXe, as well as a particle detector. 
The top and bottom of the vessel are 8" Conflat flanges. The vessel is suspended from  
a vertical 3" OD stainless steel tube, 6" in length, which penetrates its top flange. At its 
upper end, the 3" tube is welded to a 4~5/8" Conflat flange and a large stainless steel plate
which provides mechanical support. The plate also acts as part of the cryostat.
The 4~5/8" flange and tube allows direct access to the interior of
the xenon vessel from the laboratory.  With a glass viewport attached to this flange, 
the LXe can be seen inside the vessel.  This access port can also be used to 
introduce detectors, materials, or radioactive sources.  A smaller, fused-silica viewport
is welded into the bottom of the vessel; it is paired with an identical viewport in the cryostat
to admit laser light for use in LXe purity tests.

The vessel has three plumbing connections for xenon flow: a 1/4" SS liquid supply
line that enters the vessel at the top, a 3/8" liquid return line that drains the
vessel from the bottom, and a 1/4" gas return line that exits the vessel at the top;
see Figures~\ref{fig:Schematic} and \ref{fig:Vessel}.
 LXe from the condenser runs down the liquid
supply line and collects in the bottom of the vessel and in the liquid return line. 
A heater on the liquid return line is used to boil the liquid for re-circulation or
for recovery.  The gas return line at the top of the vessel allows gas to circulate  
freely through the vessel, especially when the vessel is filled with liquid. As
discussed in Section~\ref{sec:gasflow} below, free recirculation of gas is important  
for system operation. All three plumbing lines have VCR fittings so they can be  
disconnected from the rest of the system, thus allowing the condenser to be removed  
for servicing.

\subsection{Recirculation Pump}

An external, custom gas recirculation pump is used to force xenon flow through the
condenser and the xenon vessel.  It can achieve controllable xenon flow rates of
up to 10~SLPM.   The pump is a bellows-type, made entirely of stainless steel,
except for a teflon sleeve.  The pump is driven by a
1/3-HP, three-phase motor \cite{motor}; the motor itself is
controlled by an inverter \cite{inverter}, which allows flow control via adjustment of the repetition rate of the pump.

\subsection{Level Meter}

A capacitive level sensor is integrated into the liquid return line.  It has a co-axial 
cylindrical geometry formed
by suspending an 11.5" x 1/4" OD stainless steel tube inside the 3/8" OD
liquid return line.  The inner conductor is vented to allow the liquid to flow
unimpeded.  It is wrapped
at the top and bottom with a small amount of Kapton tape to prevent 
electrical shorting.
The sensor has a capacitance of 104~$\pm$~3~pF in
vacuum (224~$\pm$~3~pF including cabling ).  Because the level meter is in the  
liquid return line, rather than the vessel itself, it will only accurately
read the liquid level in the vessel when the pressures are equal. This can 
be ensured by keeping the gas return valve in Figure~\ref{fig:Schematic}
open.

Changes in capacitance are measured with a custom circuit.  The level sensor is
in series with a resistor, forming a low-pass RC filter.   
An AC voltage of amplitude 0.15~V and frequency 8~kHz is input to the filter,  
and the voltage across the capacitor is amplified and rectified for computer readout by
a data acquisition board \cite{ni}.  To maximize  sensitivity and dynamic range, the resistor 
in the filter was chosen such that 8~kHz is near the knee frequency.

The response of the circuit output was calibrated with a set of known capacitances.  
A quadratic fit of this data is used to interpolate for future measurements.  The
capacitance can then be converted to a liquid level measurement based on the
known dielectric constant of LXe.  Changes in capacitance as small as
1~pF can be measured, corresponding to a height sensitivity of about 1~mm.

\subsection{Alarm system}

We have constructed an alarm system to notify lab personnel in the event of a
serious system failure, such as a power outage. Three alarm conditions
are considered: loss of electrical power to the cryocooler, loss of electrical
power to the trim heater power supply and/or PID controllers, and an overpressure
alarm. Each alarm is represented by a switch which closes if the alarm condition
is present. The switch then activates a commerical pager unit \cite{zetron},
which dials a list of phone numbers until the alarm is acknowledged. The pager
unit derives power from a UPS to ensure that it remains active in the event that
electrical power is lost throughout the lab. Test alarms can be generated with a
push button to confirm that the system is active.

\section{The Effects of Gas Flow} 
\label{sec:gasflow}

During normal operation, the recirculation pump is used to force gas into the  
condenser, where some fraction of it condenses and travels with the remaining gas  
flow down into the vessel.  The liquid either cools the vessel through evaporation
or collects in the vessel, depending on the vessel temperature.  The gas usually
exits the system through the gas return line, although it can return
through the liquid return line as well if no liquid is present. Cryo-pumping can also
be used as a source of gas flow, but this is less convenient due to the consumption of
liquid nitrogen. 

We find that forced gas flow significantly increases the cooling power delivered  
from the condenser to the vessel. One possible explanation for this effect is that
gas flow is necessary to transport the xenon ``dew" from the cold surfaces 
of the condenser to the remote storage vessel. Regardless of the origin of the
effect, however, the importance of gas flow for cooling purposes 
magnifies the role of the gas return line in the system.  During the initial cool down, 
it makes no difference if the gas return valve is open or closed, because the liquid return 
line provides an alternate return path. But once liquid has collected, the valve must 
remain open, or else the liquid will block the flow of gas, preventing cooling power
from being delivered to the vessel.

\begin{figure}[t!]\centering
\includegraphics[width = 140 mm]{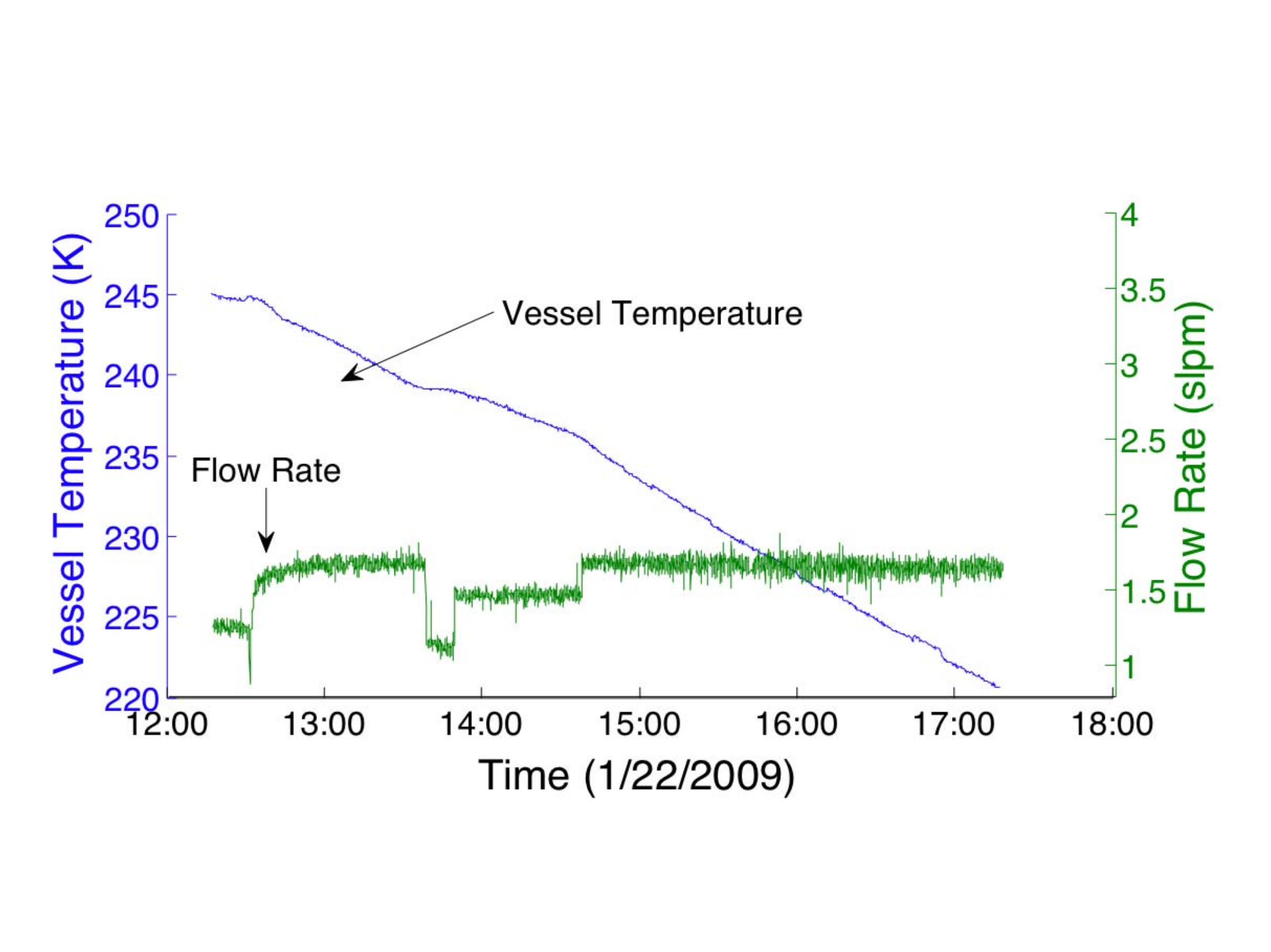}
\caption{A comparison of the xenon vessel temperature and the gas recirculation
rate during a cool-down.  Note that the temperature decreases more quickly when
the flow rate is increased.}
\label{fig:Temp_v_Flow}
\end{figure}


\subsection{Gas flow during vessel cooldown}

Starting with the system at room temperature and under vacuum, i.e.,
with no xenon gas present,
the vessel temperature drops by only a few degrees when
the cryocooler is activated, demonstrating that conductive cooling through
the SS plumbing is negligible. Once xenon gas is introduced, the xenon vessel
cools through convective heat exchange with the condenser, but the vessel
temperature levels out at roughly 200~K in the absence of forced gas flow. We
find that a minimum gas flow rate of about 1.2~SLPM is typically necessary to
cool the xenon vessel from room temperature to 
the saturation temperature. 
At higher flow rates, from 2 to 4~SLPM, the vessel cools more quickly,
as shown in Figure~\ref{fig:Temp_v_Flow}.  The effects of increasing the flow
rate even further are unclear.  In one instance, increasing the flow rate to 5~SLPM
decreased the vessel cooling rate, since the large amount of incoming warm gas heated
the condenser above LXe temperature. However, in another instance, flows of 6.8~SLPM led to greater cooling rates in the vessel, so there may be other effects, such as 
pressure, that play a role. Cooling rates are discussed further in Section~\ref{sec:coolingrates}.   

Forced gas flow may only be critical during the initial phase of vessel cooling. 
During one cool-down, gas flow and condensation were
established, but the recirculation pump was unexpectedly halted, interrupting flow
for $\sim$1~hour.  Condensation continued during this time.  Upon restarting the
pump, the xenon gas pressure rose and would not stabilize until the pump was shut
off again.  However, condensation continued and the liquid level in the
vessel rose without the forced flow.  It is therefore not clear if the 1.2~SLPM
flow rate is an absolute limit for vessel cooling; it may only be required to
initialize condensation.  

\subsection{Gas flow during liquid maintenance}
\label{sec:gasflowliquidmaintenance} 

Once the vessel is cold and filled with liquid, it is easy to maintain at constant
temperature and pressure for indefinite periods of time by forcing gas flow
through the condenser and liquid vessel, using the gas return line as the exhaust.
In this arrangement the gas circulates in a closed loop. This is generally our default 
configuration during liquid maintenance, and it allows for continuous
purification of the xenon gas with a gas phase purifier. However,
we have also studied the possibility of maintaining the liquid at constant temperature
and pressure with the recirculation pump turned off. This situation could be
important, for example, if the pump were to lose power unexpectedly. As
detailed below, we find that it is possible, but more difficult, to achieve
stability with the recirculation pump turned off.

Note that the xenon will thermodynamically recirculate at a small rate
(a few SCCM) if the external gas plumbing allows gas to flow from the system
output to the condenser input. This ``thermal recirculation" is driven by the system
heat leak, and it can be augmented by warming the liquid return line with its heater.

We achieve temperature and pressure stability most easily in the absence of  
forced gas flow by closing a valve in the external plumbing which prevents
thermal recirculation. Under these conditions, we expect that no gas will
enter the condenser input from the external system. Our flowmeter
confirms this expectation. Nevertheless, we see through the large viewport
that a steady stream of liquid drops falls into the vessel from above.  This 
indicates that gas is counter-flowing up the liquid supply line from the vessel,
liquefying, and falling back down, creating a closed heat exchange loop.
This behavior is rather sensitive to the condenser temperature. For example,
raising the pre-cooler temperature from 179.5~K to 180.5~K is enough
to disturb the establishment of this heat exchange loop.

If we turn off the external recirculation pump without preventing
thermal recirculation, then we usually find that condensing
slows or stops, and that the system temperature and pressure slowly
rise. Since the previous tests show that the system is able to condense the
cold counter-flowing gas from the LXe vessel, this behavior could indicate
that the extra heat load from the room temperature gas is too large.
It is possible that this situation could be
remedied by improving the cooling power of the condenser, particularly
that of the post-cooler. Also note that if we encourage thermal recirculation
with the heater, this makes the situation worse, increasing the rate
of temperature and pressure rise.

On the other hand, if we establish the internal heat exchange loop by
preventing external thermal recirculation, and we allow the system to run in  
this mode for several hours, we find that opening the external valve to  
allow thermal recirculation does not disturb the system stability.  That is, 
the system behaves as if the valve were still closed.
 
\section{Operation}

The operation of the system is divided into four stages: preparation,
vessel cool-down and filling, liquid maintenance, and recovery.

\subsection{Preparation}

To remove impurities in the system, the condenser and vessel are evacuated
to $\sim$10$^{-7}$~torr using a turbomolecular pump backed by a dry scroll
pump.  The cryostat is evacuated to $\sim$10$^{-3}$~torr using a similar
configuration, and it is pumped continuously
while the cryocooler is running.  A xenon gas supply bottle with a regulator 
is opened, and the regulator is set to introduce gas at a pressure 
between 1000 and 1400~torr.  The recirculation pump is turned on and set to a
flow rate of at least 1.2~SLPM.  The gas circulates though the condenser,
vessel, and external plumbing, and optionally through a zirconium getter
for purification \cite{saes}. The
pressure regulator on the xenon gas supply bottle is left open to allow
additional gas to enter the system as needed. The pre-cooler set point temperature
is set to 179.5~K.  Once this is done, the system is prepared for cool-down.

\subsection{Vessel cool-down and filling}

\begin{figure}[t!]\centering
\includegraphics[width = 140 mm]{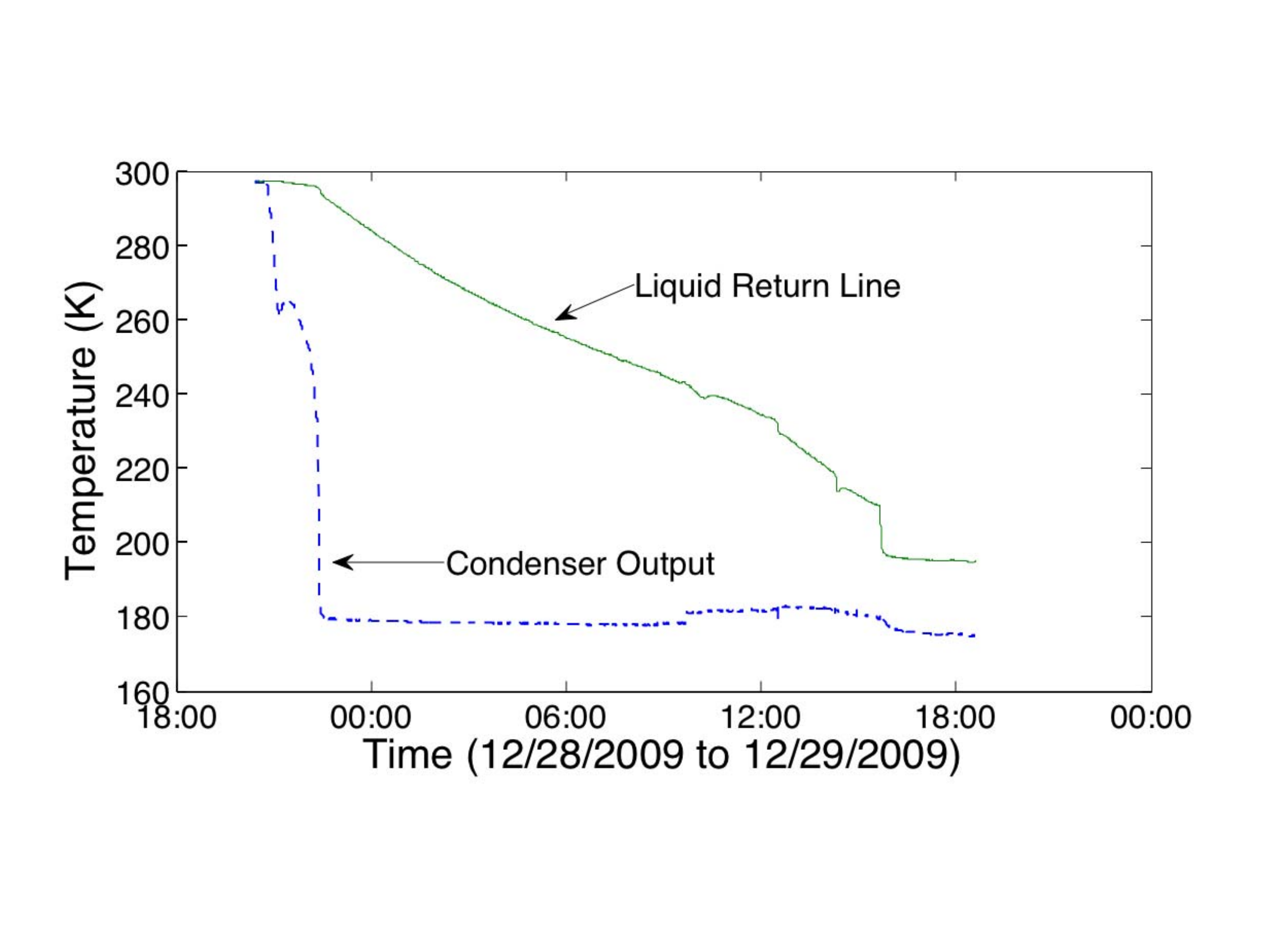}
\caption{Temperature of the output of the condenser and temperature of the
liquid return line of the vessel during a typical cool-down.  The condenser output
temperature line is kinked at about 260~K when the heaters turn on.  Other
kinks are due to intentional changes in flow rate.}
\label{fig:CoolDown}
\end{figure}


During the vessel cool-down and filling phase, the gas continues
to circulate through the condenser, vessel, and external plumbing. As the
gas cools and becomes more dense, and later as the gas liquefies, 
additional gas is delivered from the supply bottle to the system as needed to
maintain a constant pressure. The total amount of xenon in the liquid 
system can be monitored by measuring the remaining pressure in the supply bottle.

Figure~\ref{fig:CoolDown} shows a temperature history of a typical
cooldown as recorded by thermocouples at the output of the condenser and at the liquid
return line near the bottom of the vessel. The cryocooler is activated
to begin cooling the system, and the condenser temperature immediately
drops.  When the temperature reaches 260~K, the trim heaters turn on  
for the first time, momentarily warming the condenser. 
(The heaters are programmed to stay off when the condenser is above this temperature.
This avoids overheating the cold head).  After $\sim$1~hour, the temperature
of the output of the condenser falls sharply, indicating LXe has
formed and is pouring down into the vessel.  After this sharp temperature decrease, 
drops of LXe can be seen via the viewport falling onto the bottom of the vessel 
and boiling away, and the temperature of the LXe return line decreases at about six times the
previous rate. Note that this indicates that most of the cooling power is
delivered to the vessel in the form of liquid xenon, rather than gaseous xenon.

The vessel cools this way for some time, typically about 8~hours, but it can be
more or less depending on flow and pressure conditions.  When
the vessel temperature is low enough, liquid begins to collect in the bottom
of the vessel.  When enough liquid has collected, it overflows a small lip
at the drain of the vessel that leads to the liquid return line.  This
``splash'' brings the cold liquid into direct contact with the return line,
causing the temperature there to quickly drop, as can be seen in Figure~\ref{fig:CoolDown}
around 16:00 on 12/29.  This drop is correlated with a
quick rise in the measured liquid level as the level meter is filled for
the first time.

After the ``splash'' of liquid fills the return line, LXe continues
to collect in the vessel.  At this stage, the gas return bypass valve must be
open to ensure that a high rate of gas flow can continue. (We typically 
keep this valve open through the entire process, from initial preparation  
through xenon recovery.) Condensing 1~kg 
of xenon takes 2-3~hours, depending on pressure and flow conditions.

\subsection{Liquid Maintenance}

Liquid can be maintained indefinitely in the vessel at constant temperature and
pressure by keeping the pre-cooler set point temperature at 179.5~K, and by 
maintaining a nominal gas flow rate of 1-2~SLPM with the recirculation pump. 
As described in Section~\ref{sec:gasflowliquidmaintenance}, the recirculation  
pump can also be turned off under certain conditions. 

\subsection{Xenon Recovery}

The xenon is recovered by cryopumping.  The room temperature gas supply
bottle, which is made of aluminum, is placed in a liquid nitrogen bath,
freezing any xenon inside.  The supply bottle pressure regulator is fully 
opened to allow xenon gas to flow backwards through it.  In 
addition, a bypass valve which is connected in parallel with the pressure 
regulator can be opened to reduce the pumping impedance further, but this is  
usually not necessary.  

Finally the cryocooler is turned off and the LXe is 
allowed to warm, raising the vapor pressure. A hand valve between the 
liquid system and the cold supply bottle is used to manually regulate the 
gas flow rate. Care should be taken to avoid forming xenon ice by pumping  
too quickly. Recovery can be made quicker by heating the liquid return line directly with 
its integrated heater, or by filling the cryostat insulating vacuum with a dry
gas such as nitrogen or helium to provide a thermal connection to the room
temperature lab.

\section{Measurements}

\subsection{Liquid Level}

\begin{figure}[t!]\centering
\includegraphics[width = 140 mm]{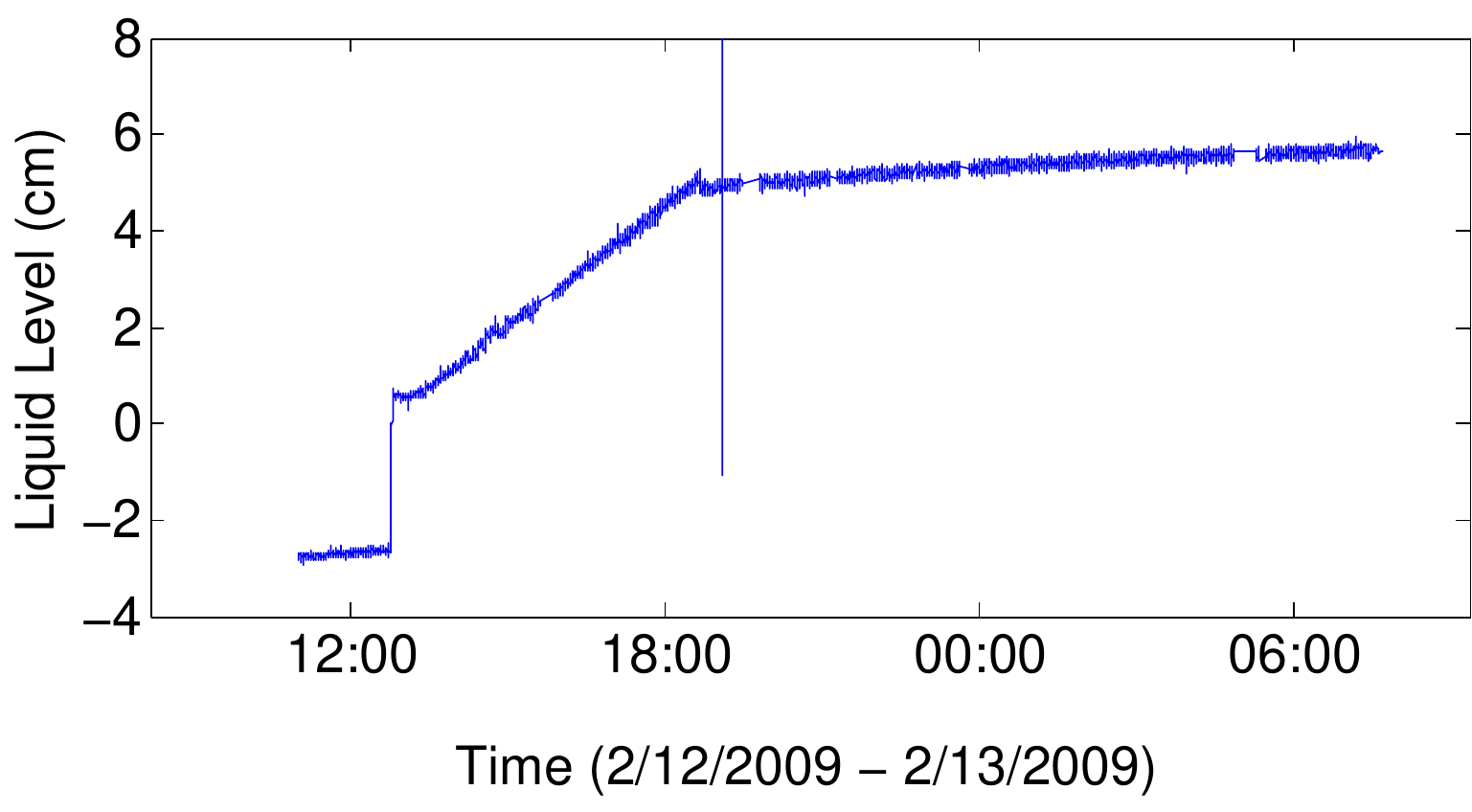}
\caption{A time history of the LXe level during condensation. The fast rise
from negative to positive values around 14:00 is due to the filling of
the liquid return line, while the slower rise over the next several hours is the
filling of the storage vessel. Note that the spike around 19:00 is spurious.}
\label{fig:Level}
\end{figure}

Figure~\ref{fig:Level} shows a sample plot of the height of LXe in
the vessel during a cool-down.  The gas pressure was 1000~torr.  The
recirculation pump forced a constant flow rate of 2.8~SLPM from 11:00
to 18:45 PM, after which it was turned off, and the xenon supply bottle
was valved off.  The initial rapid rise from -2.7~cm to 0.6~cm corresponds to the
``splash'' when liquid first overflows from the vessel into the return line. 
After that, a roughly linear increase in the height of the xenon can be seen
as the storage vessel fills with liquid. A surprising feature is that from
6:45~PM until 9:00~AM the next day, the liquid level is seen to be rising
slowly. This could indicate that the level sensor circuit has a slowly drifting
systematic error.

\subsection{Cooling Rate of the Vessel}
\label{sec:coolingrates}

The cooling rate of the stainless steel vessel was measured on two separate
trials.  During the first run, the flow rate was maintained at about 1.4~SLPM, with
a xenon pressure of 1500~torr.  The vessel cooled at 5.6~K/hr.  During another
run, the vessel cooled at 6.5~K/hr, with a much higher gas flow rate of 6.8~SLPM
and pressure of 1500~torr.  Using a specific heat for stainless steel of 500~J/kg/K,
 and estimating the vessel weight at 20~kg, we find that
these cooling rates correspond to 15.5~W and 18.9~W of cooling power
transferred to the vessel.

\subsection{Condensation Rate}

Two trials are presented in which the condensation rate is measured at various
pressure and flow parameters.  Condensation rate is measured during vessel
filling in two ways.  1.) The mass of xenon remaining in the gas supply
bottle is calculated from the known volume of the bottle, the measured pressure,
and the density of xenon gas at that pressure.  (The density of xenon gas has a
non-linear relationship to the pressure at typical bottle pressures, and this
dependence must be taken into account.)  The rate of decrease of source mass was
then calculated using central differencing and equated to the rate of condensation
in the liquid system.  2.)  The height of xenon in the vessel is measured, and
converted into a volume and mass using the known cylindrical geometry of the
system and the density of LXe.  The rate of change in liquid height then
equates to the condensation rate.  These two methods give results in good agreement
with each other, although the liquid height method is complicated by the volume
displacement of the irregularly shaped particle detector in the LXe vessel. 
In the following, we quote results based on the bottle pressure method.

In the first trial, the xenon gas pressure was set to 1550~torr, and the gas flow
rate was varied.  The condensation rates were 0.54~kg/hr for a flow of 2.65~SLPM
and 0.61~kg/hr for a flow 3.89~SLPM. In a second trial, the gas pressure was set
to 1000~torr.  The recirculation flow rate was set to 2.8~SLPM for the bulk of
the trial.  Condensation rates between 0.36 and 0.40~kg/hr were observed.

The largest rate of condensation, 0.61~kg/hr, was observed at a high
pressure and large recirculation rate, 3.89~SLPM and 1550~torr.  This implies
that 44\% of the circulating gas is condensed in a single pass.  A condensation
fraction of 58\% was also achieved, but only at the cost of a lower condensation
rate of 0.54~kg/hr.  Tests at a lower pressure of 1000~torr indicate both lower
condensation rate, $\sim$ 0.36~kg/hr, and a lower efficiency of 36\%.  Thus, the 
condensation fraction depends moderately on pressure and flow.

Using the latent heat of vaporization of xenon, 12.64~kJ/mol, and the specific heat
of xenon gas, 20.8~J/mol/K, we can calculate the cooling power implied by our
condensing rates.  For the largest condensation rate, 0.61~kg/hr, we find a cooling
requirement of 3.5~W and a latent heat removal of 16.4~W, for a total of 19.9~W. 
This is similar to the 18~W of cooling power we estimated is delivered to the stainless
steel vessel during cooling.   

\section{Discussion}
\label{sec:discussion}

We find that for reliable operation of our condenser it is essential to control the temperature
at both the top and bottom. Initial tests in which the condenser was cooled
only at the top showed that a large temperature gradient would appear along its length.
In this arrangement, the top of the condenser must be over-cooled
to allow liquefaction to occur in the lower portion of the condenser. This leaves the
condenser prone to ice formation, particularly if the gas flow rate suddenly decreases,
removing a heat source.

With a dual control mechanism, both ends of the condenser are cooled, ensuring that
temperature gradients are small. In addition, the two thermodynamic functions of the
condenser (cooling warm gas and liquefaction) are separated spatially in the condenser,
allowing for semi-independent temperature regulation with a pre-cooler and a post-cooler. 
This gives the condenser the flexibility to adjust to changing conditions, such as a change
in the gas flow rate. In practice, our post-cooler temperature is usually above
its set point temperature during condensation, and therefore its trim heater  
plays little role. Nevertheless, the additional cooling provided by the second stage of  
the cold head through the post-cooler improves temperature uniformity in the 
condenser, leading to more robust operation. To achieve greater control in the future, 
we intend to increase the cooling power of the post-cooler by reducing its thermal 
resistance.  

Xenon ice formation is a dangerous problem for external condensers such as the one
described here.  Ice can block the flow of liquid and gas to the xenon
vessel. Since the xenon flow is the only cooling mechanism for the vessel, its
interruption can lead to a dangerous rise in system temperature and pressure. Our
design is particularly sensitive to this problem, because the helical coil of the
condenser has a cross-sectional outer diameter of only 1/4". Therefore even a small
amount of ice can lead to flow blockage. A condenser coil made from larger
diameter tubing may improve the situation, or perhaps a condenser with an
altogether different geometry may be better. For our purposes, however, the
current design has proved to be adequate with appropriate safeguards.

Gas flow is essential for transferring the cooling power from the condenser to
the xenon vessel, and a xenon gas return line from the top of the vessel greatly
improves the effectiveness. There is a minimum flow rate necessary to cool the
vessel to LXe temperature, and the condensation rate increases, albeit
only slightly, with increasing flow.  Cryopumping can serve as the method for
forcing flow, but this is an awkward process that consumes large amounts of liquid
nitrogen, and flow must be interrupted frequently to warm and cool the supply and
recovery bottles. 
Therefore we find it is very useful to have a recirculation pump.  Using the pump,
gas flow can be
easily maintained for days at a time, and with an inverter controller, the flow rate
can be dialed to a desired value for easy testing.  The main drawback to the pump is
a loss of purity: our custom pump contains a teflon sleeve which is a source of
outgassing and teflon debris in the system. These problems can be solved with
purifiers and filters, however.

\section{Conclusion}

We have described a system for condensing and storing xenon where the source 
of cooling power has been removed from the vicinity of the liquid storage vessel,
facilitating the introduction of instruments and materials to the vessel.

Condensation rates as high as 0.61~kg/hr were
achieved, after an initial cool-down period of 8-10~hours.  This corresponds to a
condensation fraction of 44\% and a cooling power of about 20~W.  Changes in 
condenser design may be able to improve the condensation fraction and cooling power.  
Our design includes  
a two-stage cooling system for improved temperature uniformity and control.  
We find that a nominal gas flow rate is important for 
delivering cooling power to the vessel, and that a dedicated gas return line  
is useful for maintaining this flow when the vessel is filled with liquid. 

\section{Acknowledgments}

We thank John Carriker for his many contributions to the system described in 
this article. This work was supported by award number 0652690 from the National 
Science Foundation.

\end{document}